\documentclass[universe,article,accept,pdftex,moreauthors]{Definitions/mdpi} 
\newcommand{\xdeleted}[1]{}
\usepackage[final]{changes}

\firstpage{1} 
\pubvolume{11}
\issuenum{6}
\articlenumber{168}
\pubyear{2025}\copyrightyear{2025}
\externaleditor{Houri Ziaeepour} 
\datereceived{16 April 2025 } 
\daterevised{17 May 2025} 
\dateaccepted{20 May 2025 } 
\datepublished{24 May 2025} 
\hreflink{https://doi.org/10.3390/ universe11060168} 

\usepackage{color}
\usepackage{changes}
\usepackage{ulem}
\def \xdeleted{\deleted}

\usepackage{graphicx}
\usepackage{subfig}  

\def \be{\begin{equation}}
\def \bea{\begin{eqnarray}}
\def \eea{\end{eqnarray}}
\def \ee{\end{equation}}

\def \om{\omega}

\def \Om{\Omega} 
\def \ome{\omega_E}
\def \omgw{\omega_{\rm gw}}
\def \fgw{f_{\rm gw}}
\def \no{\nonumber}

\def \a{\alpha}
\def \bet {\beta}
\def \g {\gamma}
\def \del {\delta}
\def \vr {\vec{r}}
\def \nl {n_{\lambda}}
\def \hf {\frac{1}{2}}
\def \qtr {\frac{1}{4}}

\def \Az {A_0}

\def \Nop {N_{\rm ops}}
\def \trh {{\tilde \rho}}
\def \hthc {h_{\rm th~coh}}
\def \hthp {h_{\rm th~prod}}

\def \tg {{\tilde g}_T}
\def \tn {{\tilde n}}

\def \tH {{\tilde H}_T}
\def \C {\mathcal{C}}
\def \S {\mathcal{S}}

\Title{A Novel Search Technique for Low-Frequency Periodic Gravitational Waves} 

\TitleCitation{A Novel Search Technique for Low-Frequency Periodic Gravitational Waves}

\Author{{Harshit Raj} 
 $^{1}$\orcidA{}, Sanjeev Dhurandhar $^{2}$\orcidB{} and  Massimo Tinto $^{3,}$*\orcidC{}}

\AuthorNames{Harshit Raj, Sanjeev Dhurandhar and  Massimo Tinto}

\isAPAStyle{%
       \AuthorCitation{Raj, H., Dhurandhar, S., \& Tinto, M.}
         }{%
        \isChicagoStyle{%
        \AuthorCitation{Raj, Harshit, Sanjeev Dhurandhar, and Massimo Tinto.}
        }{
        \AuthorCitation{Raj, H.; Dhurandhar, S.; Tinto, M.}
        }
}

\address{%
$^{1}$ \quad Indian Institute of Science Education and Research, Pashan, Pune 411008, India; {harshit.raj@students.iiserpune.ac.in} 
\\
$^{2}$ \quad Inter {University} 
Centre for Astronomy and Astrophysics, Ganeshkhind, Pune {411007,} 
 India; sanjeev@iucaa.in\\
$^{3}$ \quad Divis\~{a}o de Astrof\'{i}sica, Instituto Nacional de Pesquisas Espaciais, S. J. Campos 12227-010, {SP,} 
 Brazil}
\corres{Correspondence: massimo.tinto@gmail.com}

\abstract{We quantify the advantages of a recently proposed data processing
  technique to search for continuous gravitational wave (GW) signals from
  isolated rotating asymmetric neutron stars in data measured by
  ground-based GW interferometers.  This technique
  relies on the symmetry of the motion around the Sun of an
  Earth-bound gravitational wave interferometer. By multiplying the
  measured data time series with a half-year time-shifted copy of it, we obtain two advantages: (i)
  the main Doppler phase modulation of a monochromatic gravitational
  wave signal is exactly removed, and (ii) the signal in the product data are located at \xdeleted{about} twice the GW signal frequency. The first significantly reduces the size
  of the signal’s parameter space over which a search is to be performed. The second is advantageous at low frequencies; we find that, {with currently available computer processing speeds}, this technique is capable of achieving sensitivity that is comparable to or even better than coherent and other possibly non-coherent methods. Further, since our proposed method is implemented over a year-long
  data segment, it requires processing time comparable to the data
  acquisition time of currently available computers.} 



\begin{document}

\section{Introduction}
\label{SecI}

The first observation of a gravitational wave (GW) signal made by the
Laser Interferometer Gravitational Wave Observatory (LIGO) project
~\cite{LIGO} announced on 11 February 2016~\cite{GW150914} marks the dawn of GW
astronomy~\cite{Thorne1987}. By~comparing strain data recorded by two
interferometers at Hanford (Washington) and Livingston (Louisiana),
scientists reached a high level of detection confidence that the GW
source of the observed signal was a coalescing binary system
containing two black holes of masses $M_1 = 36^{+5}_{-4} \ M_\odot$
and $M_2 = 29^{+4}_{-4} \ M_\odot$ at a luminosity distance of
$410^{+160}_{-180}$ Mpc, corresponding to a red-shift of
$z = 0.09^{+0.03}_{-0.04}$.{\endnote{{The uncertainties quoted are at
  the $90$ percent confidence level.}}
}

The two LIGO detectors constrained the direction of this binary system
to a broad region of the sky because the Italian--French VIRGO
interferometer~\cite{VIRGO} was not observing at the time of detection,
and no electromagnetic counterparts could be uniquely associated with
the observed signal. Ground-based observations inherently require the
use of multiple detectors that are widely separated on Earth and
operate in coincidence. A~network of GW interferometers operating at
the same time can (i) discriminate a GW signal from random noise
fluctuations and (ii) infer the parameters characterizing the wave's
astrophysical source (such as its sky location, luminosity distance,
mass(es), and dynamic time scale)\mbox{
\cite{Thorne1987,SchutzTinto,GT,SathyaSanjeev,JaranowskiKrolak}}.

Since the first GW observation, many other signals have been detected
by LIGO and VIRGO~\cite{LV2,LV3,LV4,LVD}, including observations of
coalescing binary systems with two neutron stars~\cite{LV5}. The~first
observation of such a signal made by the LIGO--VIRGO network was also
corroborated by gamma-ray observations~\cite{GammaRay}.  A~network of
three interferometers (such as the LIGO--VIRGO network) can 
infer the location of the source in the sky~\cite{GT}, because~it can estimate two independent time delays and rely on the asymmetry of the
detector's antenna patterns with regard to the plane containing the three
detectors' locations. The~two independent time delays provide the
location of two points in the sky where the signal may have come
from. These two points are the mirror image of each other with respect to the
plane passing through the locations of the three detectors. Since the
detector antenna patterns are not symmetric with respect to this plane, the~two-point degeneracy from the time delays can be resolved, and~the correct
location of the GW source can be obtained. In~the recent past, the~Japanese detector KAGRA~\cite{KAGRA} has also joined the network, which will further help to pin down the location. A~detector out of the plane of three detectors uniquely resolves the degeneracy of two antipodal points obtained purely by time~delays. 

The searches for monochromatic GW signals conducted so far by both
projects were only able to identify upper limits for the amplitudes that
characterize them~\cite{LSC1,LSC2,LVD}. Sources of monochromatic GW
signals, emitting in the frequency bands accessible by ground-based
interferometers, are expected to be rapidly spinning nonaxisymmetric neutron stars (such as millisecond pulsars). Although~the physical mechanisms that generate substantial nonaxisymmetry
are well understood, and~the resulting estimated gravitational wave
amplitudes have given confidence in their detections
\cite{Prix2006,SathyaBernard2009,Owen2005,Johnson-McDanielOwen2013},
it seems that to achieve a successful observation, further improvements
in detector sensitivities, together with additional data processing techniques, might be~required.

{Since their early designs, ground-based GW interferometers have attempted to achieve better sensitivities at low frequencies. This is because the number of GW signals increases in this part of the GW band, as do their amplitudes. The~latter statement follows directly from the theory of general relativity, as~the energy radiated by any GW source is proportional to the product of the square of the amplitude and the square of the frequency of the radiated signal. If, for~example, we consider two similar GW sources emitting the same amount of energy but at different frequencies, the~one emitting at a lower frequency would be characterized by a GW signal of larger amplitude. Detection in this frequency region would shed light on the physical structure and equation of state of spinning compact stars.}

Here, we investigate the advantages of a recently proposed data
processing technique~\cite{TintoCW2021} to minimize the size of the
parameter space that characterizes continuous GW signals and to
improve the sensitivity of the resulting GW searches {at low frequencies}. The~technique we
discuss is coherent and is implemented on a dataset that is equal to
the product of the original data with a time-shifted (about
half a year) copy of it. The~resulting GW signal in the new dataset is
characterized by a phase that is, in~principle, unaffected by the
Doppler modulation and has a carrier frequency that is twice that of
the GW. Since the GW frequency is higher than the original GW frequency, at~low frequencies, we are at an advantage in two ways: (i) GWs with frequency below the lower end of the bandwidth of the detector become accessible, because~the product signal lies in the bandwidth if the GW frequency is greater than \xdeleted{the} half the low-frequency cut-off. {If, let us say,
we assume the lower frequency cut-off at $10$ Hz, then signals
with frequencies falling in the interval (5{--}
10) Hz would become
detectable by our method. This is the trump card of our method.} (ii) Since the noise curve has a negative slope at low frequencies, this technique is clearly advantageous. As~a result, we quantify its ability to search for continuous GW signals at these frequencies. We find that, in~this frequency
region, our technique is faster and more sensitive than a coherent
search. We could not compare our technique with semi-coherent ones,
as there are no published results in this frequency regime.
\xdeleted{It should be emphasized that, because~of the doubling of the GW
frequency in the combined dataset, this data processing technique
allows us to search for continuous signals at frequencies lower than
the interferometer's noise spectrum frequency cut-off.} 

An outline of our article is provided below. In~Section~\ref{paradigm}, we
present a summary of the mathematical properties of the data
processing technique on which we will focus~\cite{TintoCW2021}. To~provide a general and simple framework for its understanding, here we
assume the Earth's trajectory around the solar system barycenter (SSB)
to be perfectly circular and characterized by a period exactly equal
to $365$ days. After~showing that the phase modulation of a sinusoidal
gravitational wave signal measured by an interferometer is equal in
magnitude and opposite in sign for samples that are six months apart, 
we demonstrate that the product of a year-long dataset with a copy of
itself that is time-shifted by six months results in a new dataset in
which the phase of the resulting GW signal no longer depends on the
signal's source sky location. In~Section~\ref{SecIII}, we further
investigate the circular trajectory for which the orbital period
is now the exact one-year time-lapse, and we include the effects due to the
pulsar frequency spin-down. Since the Doppler modulation due to the
motion around the SSB still cancels out, we are now left with the
residual Doppler modulation due to the Earth's diurnal rotation.  To~quantify the gains offered by this non-linear technique over full
coherent searches and other proposed techniques, we analyze
the statistical properties of the noise entering the newly synthesized
dataset and estimate the number of patches in the parameter space
over which a search should be performed. We find that the number of patches needed to implement our method is typically $10^4$ times smaller than the coherent search. This reduces the computational cost of our search by the same factor, in addition to reducing false alarms. We show that for the O4 sensitivity curve, our method performs better than the coherent method for frequencies of $\lesssim$17 Hz. Furthermore, our method allows one to search for frequencies $\leq$10 Hz, which are not observable by other~methods.

\section{The~Paradigm}
\label{paradigm}

Millisecond pulsars are the typical sources of continuous, narrowband, quasi-sinusoidal GW signals searched by ground-based GW
detectors. The~space--time quadrupolar perturbations they emit can be
modeled by a family of parametrized template waveforms. Because~of the
long observation times required to achieve their detection, one must
include in the modeling the temporal changes in the wave's frequency
due to the pulsar's spin-down effect and the motion of the detector
relative to the SSB (which can be regarded as an inertial reference
frame). An~accurate model of the GW signal measured by a detector is
required to effectively account for these frequency modulations in the
analysis of data and maximize the likelihood of detection. A~set
of parameters defines the instantaneous phase of the signal measured at
the detector, $\Phi (t)$, and~the corresponding instantaneous
frequency, $f(t) \equiv \frac{1}{2 \pi} \ \frac{d \Phi}{dt}$.  These
can be written in the following {forms}
~\cite{BradyCreighton1998}:
\begin{eqnarray}
\Phi (t) & = & \Phi_0 + 2 \pi f_0 \left[t + \frac{\vec{r} (t)\cdot \hat{n}}{c} + 
                 \frac{f_1}{2} \left(t + \frac{\vec{r} (t) \cdot \hat{n}}{c}\right)^2\right] \ ,
\\
\frac{1}{2\pi} \ \frac{d \Phi(t)}{dt} \equiv f(t) & = & f_0 \ \left(1 + \frac{\vec{v} (t)\cdot \hat{n}}{c}\right) \ \left[1 + f_1
             \left(t + \frac{\vec{r} (t)\cdot \hat{n}}{c}\right)\right] \ ,
\label{FreqPhase}
\end{eqnarray}
where $f_0$ is the signal frequency in the rest frame of the pulsar,
$f_1 = {\dot f}/f_0$ is the first pulsar's frequency spin-down term,{\endnote{{Our
  analysis will only include the first derivative of the variation in
  the pulsar's angular velocity}.}} $\hat{n}$ is the unit vector to the
pulsar's sky position relative to the SSB, ($\vec{r}(t)$
$\vec{v} (t)$) are the position and velocity at time $t$ of the
detector relative to the SSB, respectively, and~$c$ is the speed of
light in the vacuum.  The~spin-down parameter $f_1$ can take values in the
interval $|f_1| < 1/\beta$, where $\beta$ is the characteristic time
scale over which the fractional relative change in the pulsar spin is
of the order of unity.  If~($\a, \del$) are the right ascension and declination of the pulsar's sky location, the~set of parameters (in principle unknown) characterizing the above two functions and over
which a search is performed can be represented by the following {vector}: 
${\pmb \lambda} \equiv (f_0, f_1, \a, \del)$.

The GW strain measured at the interferometer, $s (t)$, is a linear
combination of the wave's two polarization components, resulting in the
following form:
\begin{equation}
s (t, \vec{\lambda}) = A(t) \ \cos[\Phi(t, \pmb{\lambda}) + \Psi_0 ] \ .
\end{equation}

Here, for~simplicity, we will regard the amplitude $A$ and phase
$\Psi_0$ (due to the detector's beam-pattern functions through its
orientation to the source) as constant. Although~this is a fairly
accurate assumption, as these two functions change very slowly over the
course of a day~\cite{JaranowskiKrolakSchutz1998,CutlerSchutz2005},
our simulations presented later on will include their exact time~dependences.

In~\cite{TintoCW2021}, it was first shown how to take advantage of the
symmetry of the motion of the interferometer as it rotates around the
Sun and simultaneously around the Earth's axis. Here, we will assume
that the Earth will follow a perfectly circular trajectory around the SSB
with a period equal to $365$ days. For~the sake of describing the
idea, disregard the pulsar spin-down, $f_1$. In~Section {\ref{subsec_case3}},
we will demonstrate how to account for the exact value of the year, and in
Section {\ref{el_orb}}, how to compensate for the elliptical trajectory
so that the non-linear data processing technique described here can
still be~implemented.

Since, at time $t$, the~velocity of the interferometer with respect to the SSB is
opposite in sign but equal in magnitude to the value it acquires at a
later time $t + T_0$ (with $T_0$$\sim$${\rm 6 \ months}$), and~because the
position of the interferometer at time $t+T_0$, ${\vec r} (t+T_0)$ is
equal to $-\vec{r} (t) + \vec{\xi}$ (with $\vec \xi$ being a constant
vector\xdeleted{of integration}). {The vector $\vec{\xi}$ is explained in Section~\ref{SecIII} below Equation~(\ref{eq_13}).} The expression of the detector's GW response
at these two times can be written as follows:
\begin{eqnarray}
  s (t) & = & A \ \cos\left[ 2 \pi f_0 [t + \frac{\vec{r} (t)\cdot \hat{n}}{c}] + \Psi_0 \right],
\label{Rt}
  \\
  s (t + T_0) & = & A \ \cos\left[2 \pi f_0 [t + T_0 - \frac{\vec{r} (t)\cdot
                  \hat{n}}{c}] + \Psi_0 + \rho_0 \right] \ ,
\label{RtT}
\end{eqnarray}
where we have denoted with $\rho_0$ the constant associated with the
integrations of the anti-symmetric condition fulfilled by the velocity
vector at times $t$ and $t + T_0$. If~we now multiply the two data
streams, after~some simple algebra, we find the following expression
for the GW signal in the resulting new time series:
\begin{eqnarray}
Q(t) &\equiv& s (t) s (t + T_0)
\nonumber
\\
& = & \frac{A^2}{2} \ \left[\cos(2
    \pi \ f_0 (2t + T_0) + 2 \Psi_0 + \rho_0) + \cos(4 \pi f_0 \frac{\vec{r} (t)\cdot \hat{n}}{c} - T_0 - \rho_0)\right] \ .
  \label{product}
\end{eqnarray}

{The} 
 quantity $Q$, which we term the quadratic or product signal, is equal to the sum of two terms in the new time series, one with 
a frequency of $2f_0$ and the other with a frequency close to zero. We are interested in the first term, which in the simplest model (Case I) does not depend on $\a, \del$. However, the~amplitude $A$ is a slowly varying function of time on the timescale of a day. So, essentially, a~Fourier transform of the time series is sufficient with some additional operations (see Section \ref{subsection:I} for the statistics in case I). In~practice, the~sky location of the source, the~incommensurate rotational and orbital periods of the Earth, and~the non-negligible spin-down $f_1$ lead to a search over a relatively large number of sky patches, but~the number is much smaller than that required in other methods. In~this situation, a~detailed and careful analysis is necessary. Finally, the~effects
of the elliptical motion of the Earth around the barycenter also require attention, and~this is presented in this~article.

\section{The Product~Signal}
\label{SecIII}

We start with the signal at the detector before computing the product signal. We work with the dominant quadropolar signal at a frequency twice that of the rotation frequency of the neutron star. See reference~\cite{JaranowskiKrolakSchutz1998} (and the papers cited therein). Then, the  signal $h(t)$ is provided by the following:
\begin{equation}
        h(t) = F_{+}(t)h_{+}(t) + F_{\times}(t)h_{\times}(t) \,,
    \end{equation}
where
\begin{equation}
        h_{+} = \frac{1}{2}h_{0} (1 + \cos^2 \iota) \cos\Psi(t),
    \end{equation}
\begin{equation}
        h_{\times} = h_{0} \cos \iota \sin\Psi(t) \,.
    \end{equation}

{Here}, $h_0$ is the amplitude of the signal, and $\iota$ is the angle between the total angular momentum vector of the star and the direction from the Earth to the star.
We consider a simple model that will last up to one spin down. Then, we can write the phase as follows: 
\begin{equation}
\label{eq_double_freq_phase}
    \Psi(t) = \Psi_0 + \omgw \bigg(t + \frac{\vr_d \cdot\hat{n}}{c} +  \hf f_1 t^2 + \frac{\vr_d \cdot \hat{n}}{c} f_1 t \bigg) \,,
\end{equation}
where $\vr_d = \vr_{ES} + \vr_{DE}$, and~$\vr_{ES}$ and $\vr_{DE}$ are position vectors connecting the Sun to Earth, and~Earth to the detector, respectively. Note that the quadratic term in the Doppler correction has been dropped. The~spin-down parameter is $f_1 = {\dot f_0}/f_0$. Also, we find it convenient to deal with circular frequencies, so we write $\omgw = 2 \pi f_0$. These notations and approximations are consistent with~\cite{JaranowskiKrolakSchutz1998,BradyCreightonCutlerSchutz1997}.

For~simplicity, we will assume a circular orbit for Earth with a radius equal to one AU. In~Section~\ref{el_orb}, we will argue how our analysis can be adapted to the elliptical case and also to the exact ephemeris. We believe that the results derived here will not differ greatly from the actual orbit of Earth. 
\par

In order to obtain insight into the problem and also fix ideas, we address the problem in three stages, with~increasing degrees of~complexity:
\begin{itemize}
\item Case I: We assume that the orbital period $2\pi/\Om$ is an odd integer multiple, namely 365, of~the rotational period $2 \pi/\ome$ of Earth, or, $\ome = 365~\Om$. Also, we assume that the neutron star does not spin down, i.e.,~its frequency remains constant throughout the observation period in the source frame, that is, we take $f_1 = 0$.
\item Case II: We relax the first condition and assume actual periods for orbital and rotational motion. They are not related by an integer multiple. Specifically, we assume \mbox{$\ome \simeq 366.25~\Om$} (note that the relevant period here is a sidereal day, not the mean solar day). However, here we also assume the spin-down to be zero, or~$f_1 = 0$.
\item Case III: We relax the above two conditions; orbital and rotational periods not related by an integer and also one spin-down are assumed. Thus, $\ome = 366.25~\Om$ and $f_1 \neq 0$.
\end{itemize}

\subsection{Case~I}
\label{subsection:I}

Let the detector be located at latitude $\theta_l$ and let its orientation be described by angle $\gamma$. Angle $\g$ is the angle measured from the local east direction to the angular bisector of the arms of the detector. In~Equation~(\ref{product}), we consider only the first term, which is at twice the GW signal frequency that is relevant. Then, the product signal $H(t; T_0)$ is of the following~form:\vspace{-12pt}
\begin{adjustwidth}{-\extralength}{0cm}
\centering 
\begin{equation}
    \begin{aligned}   
H(t;T_0) \equiv h(t) h(t+T_0) = & \sum_{k=-2}^2 A_k \cos(\Psi + \overline{\Psi} + 2 k \omega_E t)
 + B_k \sin(\Psi + \overline{\Psi} + 2 k \omega_E t) \,,
        \label{eq:gnrl_signal}
    \end{aligned}  
\end{equation}
\end{adjustwidth}
where quantities $A_k$ and $B_k$ are defined in Appendix \ref{Prod_Sig}. Phases $\Psi$ and $\overline{\Psi}$ are described by the following:
\begin{equation}
\begin{aligned}
\Psi &\equiv \Psi(t) = \Psi_0 +  \omgw t + \frac{\omgw}{c} \hat{n}.\vec{r}_d(t) \, \no \\
\overline{\Psi} \equiv & \Psi(t + T_0) 
    = \Psi_0' + \omgw t + \frac{\omgw}{c} \hat{n} \cdot \vec{r}_d (t + T_0)  \,.
\end{aligned}
\end{equation}

{In} this case, the~choice of $T_0$ is simple: it is exactly half a year or $T_0 = \pi/\Om$. In~this model, the choice of $T_0$ cancels both the orbital and rotational Doppler modulations. $\Psi_0$ and $\Psi'_0$ are the initial phases corresponding to the initial times $t$ and $t + T_0$, which are six months apart. Adding the phases results in the following expression:
\begin{equation}
                \Psi + \overline{\Psi} = \Psi_0 + \Psi_0' + 2 \omgw t + \Phi_0' = 2 \omgw t + \Phi_0 \equiv \Phi_I \,
                \label{eq:phase_0}           
\end{equation}   
where
\begin{equation}
\label{eq_13}
        \Phi_0 = \frac{\omgw}{c} (\vec{r}_d(t) + \vec{r}_d(t+T_0)) \cdot \hat{n} + \Psi_0 +  \Psi'_0 \,.
\end{equation}

Quantity $\Phi_0$ is constant because the sum of the vectors inside the brackets results in a constant vector {$\vec{\xi}$} pointing along the axis of the rotation of Earth. Thus, phase term $\Phi_I$ is essentially simple, with~a frequency twice the GW signal frequency. However, the~amplitude varies on the time scale of a day, and~then, the product signal is a Fourier series as provided by Equation~(\ref{eq:gnrl_signal}). 
\par

To fix ideas, we consider a simple, specific situation. We take the source to lie on the celestial equator, and~other relevant parameters are set as follows: $\alpha = \pi, \delta = 0$ and $h_0 = 1$ and the constant phase $\Phi_0 = 0$ (a non-zero phase $\Phi_0$ does not matter because, finally, in~the statistic, we take the modulus of the amplitudes). We also consider the detector to be located at the equator with its arms along the north and east directions. Then, the product signal is as follows:    \vspace{-12pt}
    \begin{sloppypar}
\begin{align}   
                    H(t; T_0) &= \frac{19}{64} \cos(\Phi_I 
                    + \quad+ \frac{3}{32} \left[ \cos\left(\Phi_I + 2 \ome t\right) + \cos\left((\Psi + \overline{\Psi}) - 2 \ome t\right) \right] \nonumber \\
                    &\quad+ \frac{1}{128} \left[ \cos\left((\Psi + \overline{\Psi}) + 4 \ome t\right) + \cos\left((\Psi + \overline{\Psi}) - 4 \ome t\right) \right] \,.
\label{eq:spcl_waveform}    
            \end{align}        
    \end{sloppypar}

Here, note that we have ignored the constant phase as it does not affect the statistic (Equation (\ref{eq:stat_simplecase})). We now define the statistic as a matched filter that involves finite-time Fourier transforms. The~finite-time Fourier transform of a function $g(t)$ over time interval $T$ is defined as follows:
\be
\tg (\om) = \int_{-T/2}^{T/2} ~g(t) e^{-i \om t} dt \,.
\ee

{The} integral is taken symmetrically from $- T/2$ to $T/2$. The~result is expressed in terms of sinc functions with a large argument. These are finite-time delta functions 
$\del_T (\om)$ that are approximations to delta functions introduced in Dhurandhar~et~al.~\cite{DKMW2008}. We describe their properties in Appendix \ref{append_deltafn}. For~finite times, two finite-time delta functions centered at frequencies that differ by a small amount interfere. The~interference is small if $\ome T$ is sufficiently large. The~situation is analyzed in some detail in Appendix \ref{append_deltafn}. For~the special case provided by Equation~(\ref{eq:spcl_waveform}), the~sinc terms are absent, and the Fourier transform is of the following form:\vspace{-12pt}
\bea
\tH (\om )&=& \hf [(\Az + A_1 + A_{-1}) \del_T (\om - 2 \omgw) + (\Az + A_1) \del_T (\om - 2 \omgw - 2 \ome)  \, \no \\
&+& (\Az + A_{-1}) \del_T (\om - 2\omgw + 2 \ome) \, \no \\
&+& A_1 \del_T (\om - 2\omgw - 4 \ome) ~+~A_{-1} \del_T (\om - 2\omgw + 4 \ome)] \,.
\label{eq:mf_out}
\eea

{Thus,} in the Fourier domain, one obtains  peaks at the following frequencies: $2 \omgw, 2 \omgw \pm 2 \ome, 2 \omgw \pm 4 \ome$. \textls[-15]{In~the general case, $B_k$ terms will also contribute, and~then the coefficients appearing in front of the finite-time delta functions will be complex, but~again, we will have five peaks. Given this situation, the~appropriate statistic is as~follows:}
\begin{equation}
\begin{aligned}
c_T (\om) = |\tH(\om)&|^2 +  |\tH(\om + 2 \ome)|^2  + |\tH(\om - 2 \ome)|^2 \\
          &+  |\tH(\om + 4 \ome )|^2 +  |\tH(\om - 4 \ome)|^2 \,.
\label{eq:stat_simplecase}
\end{aligned}
\end{equation}

\begin{figure}[H]
{   \captionsetup{position=bottom,justification=centering}
\subfloat[]{%
        \includegraphics[width=0.45\linewidth]{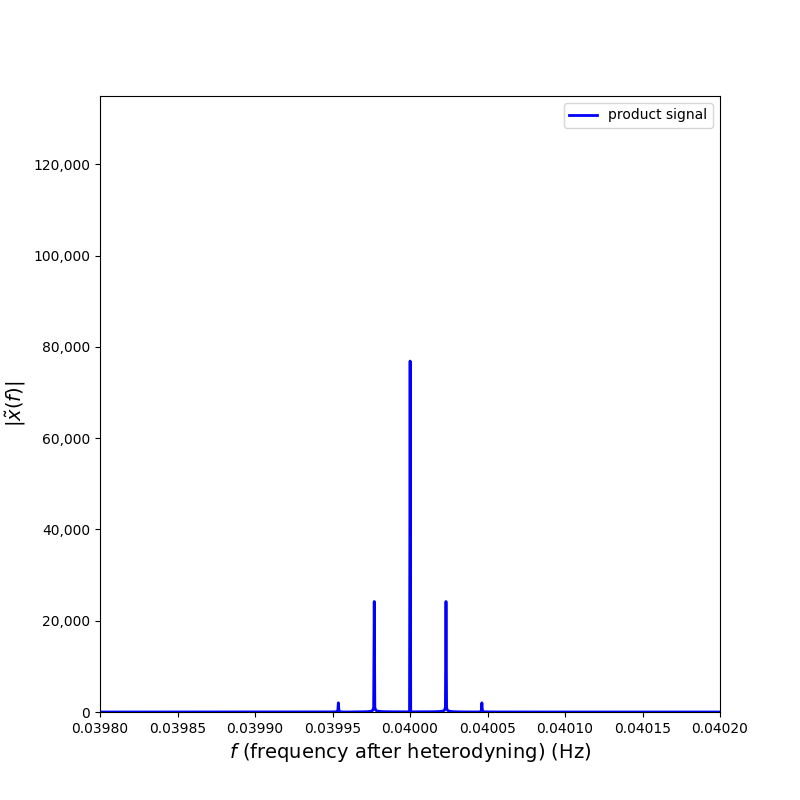} 
        \label{fig:first}
    }
    \hspace{16pt}
    \centering\subfloat[]{%
        \includegraphics[width=0.45\linewidth]{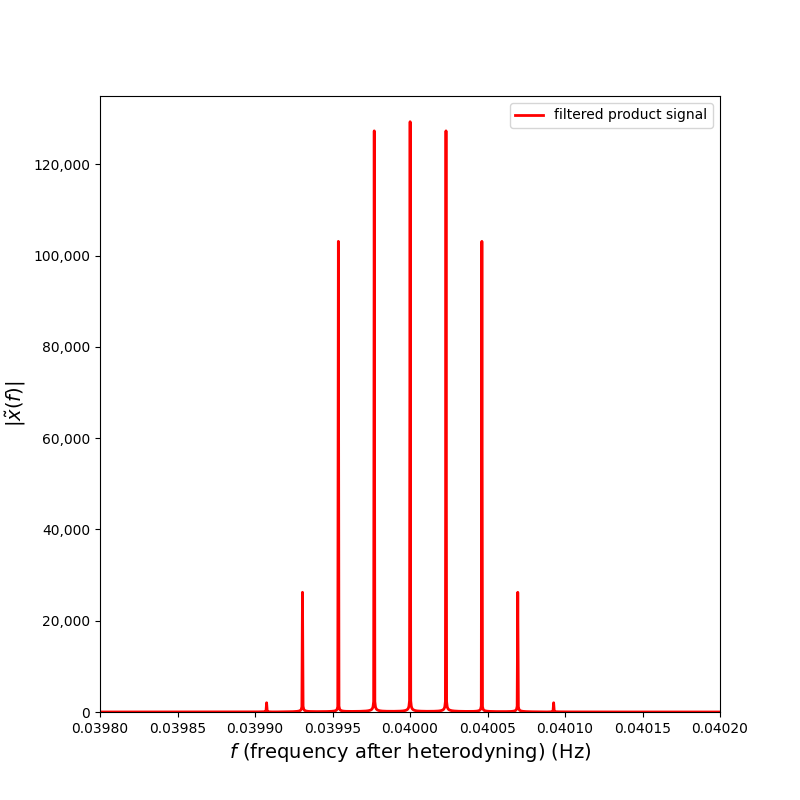} 
        \label{fig:second}
    }}
    
\caption{{Comparison} 
 of the absolute value of the DFT of the filtered and unfiltered quadratic~signals. (\textbf{{a}
}) Modulus of the DFT of the quadratic signal. (\textbf{b}) Modulus of the DFT of the filtered quadra\mbox{tic si}gnal.}
    \label{fig:main}
\end{figure}

This filter produces the same result as when operating on the signal as the matched filter. The~result is the sum of the modulus squares of all the coefficients multiplied by factor $T/2$. {This is shown in Figure~\ref{fig:main}.} Here, there are nine peaks as the filter moves along the frequency axis, but~now, the~central peak is the highest, corresponding to the matched filter; it is the peak in which all coefficients contribute. It is the peak of our~interest.
\vspace{-24pt}
In this case, just taking the Fourier transform of the time series suffices, because~the matched filter is provided by the statistic in Equation~(\ref{eq:stat_simplecase}). 

\subsection{Case~II}
\label{case II:}

We relax the assumption that the orbital period is an integral multiple of the rotational period of Earth. We, nonetheless, ignore spin-downs. But~now, the choice of $T_0$ must be made judiciously because there is no choice of $T_0$ that cancels both orbital and rotational Doppler shifts. Then, how do we choose $T_0$? The criterion we adopt minimizes the computational cost of the search. Since no choice of $T_0$ exactly cancels out Doppler modulations, the~method involves searching over sky directions or sky patches, such that over a given sky patch, one demodulation suffices. Assuming that demodulation has been performed, the~signal becomes essentially monochromatic, and~only a Fourier transform is required to extract the signal. Since the main cost of the search comes from computing Fourier transforms, this means we must minimize the number of sky patches over which the search must be carried out (the cost of demodulation can be ignored).     
\par

\subsubsection{The Choice of $T_0$ and the~Phase}

We have two choices at our disposal: we can cancel either the orbital or the rotational Doppler modulation. We choose to cancel the orbital Doppler modulation because its effect is larger and, therefore, it leads to a smaller number of sky patches. 
\par

Here, the~sidereal day is relevant. In~one sidereal day, Earth makes one full rotation of $360^\circ$ in the SSB frame. Then, there are approximately 366.25 sidereal days in a year. Thus,
\begin{equation*}
    \om_E \simeq 366.25 \times \Om.
\end{equation*}

{From} case I, the~value of $\Om$ remains the same, but~$\ome$ changes- 
 $2 \pi/\ome$ is about 4~min shorter than 24 h, which constitutes a solar day. In~this case, the~phase of the product signal is as follows (ignoring any constant phase): 
\bea
\Psi + \overline{\Psi} &=& 2 \omgw t + \frac{\omgw}{c}(\vec{r}_{DE}(t) + \vec{r}_{DE}(t+T_0) ) \cdot \hat{n}  \,, \no \\
&=&  2 \omgw t + \frac{\omgw}{c}(r_p \cos \delta \cos (\omega_E t - \alpha + \chi) ) \equiv \Phi_{II} (\omgw, \a, \del)  \,.
\label{eq:phaseII}
\eea
where $\chi$ is a constant depending on the initial configuration of the detector, independent of the signal parameters, and $r_p$ is the projection provided by the following:
\begin{equation}
\label{eq:r_p}
r_p = \sqrt{2R_E^2 (1 + \cos \omega_E T_0)}\cos\theta_l \approx 2 R_E \cos \theta_l \,.
\end{equation}

{Here}, $R_E$ is the radius of Earth, and $\theta_l$ is the latitude of the location of the detector. We find $r_p$ is of the order $10^4$ km, which is approximately the order of the diameter of Earth. If, on~the other hand, we had chosen to cancel the rotational Doppler modulation, the~corresponding distance would have been of the order of $10^5$ to $10^6$ km. Since the number of sky patches is proportional to $r_p^2$, as will be shown later in the text, we chose to cancel the orbital Doppler modulation and, therefore, chose $T_0 = \frac{\pi}{\Om}$. In~Equation~(\ref{eq:r_p}), we made this choice in order to obtain an approximate value of $r_p$. We adhere to this choice of $T_0 \simeq 1.5 \times 10^7$ s throughout the~paper.

\subsubsection{The Phase Metric and the Number of Sky~Patches}

We see from Equation~(\ref{eq:phaseII}) that because of the involved dependence of phase $\Phi_{II}$  on $\a$ and $\del$, one needs to search over directions, which results in sky patches; over a specific sky patch, the~Doppler modulation is small and below a limit (the mismatch) that we will specify. This usually results in a large number of sky patches over which the search must be carried out. The~number of patches and the size of each patch are conveniently evaluated by computing the phase metric. This metric has been provided in the literature~\cite{Dhurandhar2001}. 
We relabel the parameters $2 \omgw, \a, \del$  as $\lambda^0 = 2 \omgw, \lambda^1 = \a, \lambda^2 = \del$, which we denote collectively as $\lambda^\a$. The~metric is then provided by the following:
\be
g_{\a \bet} = \langle \Phi_{II~\a} \Phi_{II~\bet} \rangle - 
\langle \Phi_{II~\a} \rangle \langle \Phi_{II~\bet} \rangle \,,
\label{eq:metric_gen}
\ee
where the suffix, say $\a$, denotes the derivative with respect to $\lambda^\a$, and the angular brackets denote time averages defined as follows: for a function $X(t)$ defined over time interval $[0, T]$, the~time average of $X(t)$ is equal to the following expression:
\be
\langle X \rangle = \frac{1}{T} \int_0^T dt~ X(t) \,.
\ee

{The} metric $g_{\a \bet}$ includes $2 \omgw$ as a parameter over which we need to maximize. Geometrically, this amounts to projecting the metric orthogonal to $\lambda^0$. Then, the metric over the search parameters is as follows:
\be
\g_{ij} = g_{ij} - \frac{g_{0i} g_{0j}}{g_{00}} \,.
\label{eq:proj_metric}
\ee

{The} Latin indices $i, j$ comprise parameters $\a, \del$. We will take time $T$ to be $N$ number of days, where $N \gtrsim 100$.
\par
Without further ado, we compute the metrics. We find the limit of $T$ to be larger than, say, 100 days:
\be 
g_{00} = \frac{T^2}{12}, ~~g_{\a \a} = \hf \nl^2 \cos^2 \del, ~~g_{\del \del} = \hf \nl^2 \sin^2 \del \,.
\ee

Quantity $n_\lambda = \omgw r_p/c = k_{\rm gw} r_p = 2 \pi r_p/ \lambda_{\rm gw}$, where $k_{\rm gw}$ is the wavenumber, and $\lambda_{\rm gw}$ is the wavelength of the GW.  $\nl$ is essentially the number of wavelengths that fit into $2 r_p$. The~cross terms $g_{0 \a}$ and $g_{0 \del}$ do not scale with $T$, and~so when divided by $g_{00}$, which scales as $T^2$,  the~second term in Equation~(\ref{eq:proj_metric}) tends toward zero as $1/T^2$. Thus, $\g_{\a \a} \simeq g_{\a \a}$ and $\g_{\del \del} \simeq g_{\del \del}$. Also, we find that $\g_{\a \del}$ scales as $1/T$ and, therefore, tends toward zero when the observation time is a large number of days. We are now ready to write down $\g_{ij}$:
\be
\label{eq_yy}
\g_{\a \a} = \hf \nl^2 \cos^2 \del, ~~ \g_{\a \del} \approx 0, ~~ \g_{\del \del} = \hf \nl^2 \sin^2 \del \,.
\ee

{From} this, we obtain the following:
\be
\det \g = \frac{1}{4} \nl^4 \cos^2 \del \sin^2 \del \,,
\ee
and integrating $\sqrt{\det \g}$ over $\a$ and $\del$, we obtain the parameter space area $A$ as follows:
\be
A = \pi \nl^2 \,.
\label{eq:area_par}
\ee

{$A$} turns out to be approximately $\sim$1400  for $f_{\rm gw} = 100$ Hz. The~size  of each patch for the mismatch  of $\epsilon = 0.3$ is $\Delta A = \Delta l^2$$\sim$0.6. Thus, the number of patches is \mbox{$N_{\rm patch} = A/\Delta A$$\sim$2300}, and they scale $\propto f_{\rm gw}^2$. At~low frequencies, which we are interested in, $f_{gw} = 15$ Hz and $N_{\rm patch} = 51$.

\subsection{Case~III} \label{subsec_case3}

In this section, we consider signals with one spin-down parameter $f_1$. Note that considering one spin-down parameter is sufficient for signals at low frequency~\cite{BradyCreightonCutlerSchutz1997, JaranowskiKrolakSchutz1998}. 

\subsubsection{The Phase and Its Approximate~Form}

The analytical expression for the phase of the product signal (with $T_0 = \frac{\pi}{\Om}$) in terms of parameters $\boldsymbol{\vec{\lambda}}\equiv\{\lambda^0, \lambda^1, \lambda^2, \lambda^3\} \equiv \{2\omgw, \a, \bet, f_1\}$ is provided by the following (again, ignoring the constant phase):
\begin{equation}
\begin{aligned}
\label{eq:phase_approx_1}
  \Phi_{III}(t; \boldsymbol{\vec{\lambda}}) \approx~  & 2 \omega_{\text{gw}} t + \frac{\omega_{\text{gw}}}{c} r_p \cos\delta \cos(\omega_E t - \alpha + \tilde{\chi}) \\
    &+ \frac{\omega_{\text{gw}} f_1}{2} \bigg( 2t^2 + 2T_0 t + T_0^2 \\
    &+ \frac{2T_0}{c} R_o \big( \cos\alpha \cos\delta \cos\left(\chi + \Omega t\right) \\
    &+ \left( \cos\delta \cos\epsilon \sin\alpha + \sin\delta \sin\epsilon \right) \sin\left(\chi' + \Omega t \right) \big) \bigg) \,,
\end{aligned}
\end{equation}
where $\tilde{\chi}$, $\chi$, and $\chi'$ are constants independent of the signal parameters. $\epsilon$ is the tilt of Earth's axis, $R_o$ is the radius of Earth's orbit around the Sun, and~$r_p$ is provided by Equation~(\ref{eq:r_p}). This approximation is justified because the rotational Doppler terms in the expression with spin-down are negligible compared to the orbital Doppler term. Additionally, we consider $\epsilon = 0$, assuming that Earth's axis is perpendicular to its orbital plane. We expect this simplification to have an insignificant effect on the calculations of the metric. This has been tested for the case of the fully coherent method presented in~\cite{JK3}. Thus, the~expression for the phase we use for computing the metric simplifies to the following:
\begin{equation}
\begin{aligned}
\label{eq_approxprod_one_spind_phase}
  \Phi_{III} (t; \boldsymbol{\vec{\lambda}}) \approx &2 \omega_{\text{gw}} t + \frac{\omega_{\text{gw}}}{c} r_p \cos\delta \cos(\omega_E t - \alpha + \tilde{\chi}) \\
    &+ \frac{\omega_{\text{gw}} f_1}{2} \bigg( 2t^2 + 2T_0 t + T_0^2 + \frac{2T_0}{c} R_o \cos\delta \cos(\Omega t - \alpha + \chi')\bigg) \,.
\end{aligned}
\end{equation}

\subsubsection{The Phase Metric and the Number of Sky~Patches}

As can be seen from Equation~(\ref{eq_approxprod_one_spind_phase}), the~phase of the product signal is fairly complicated. To~obtain the analytical results, we further simplify this for a special regime and state numerical results for the general case.
We observe that the terms with $r_p$ and $R_o$ in Equation~(\ref{eq_approxprod_one_spind_phase}) are comparable. We seek a regime in which we can ignore the term in $R_o$, thus simplifying the analytical calculations. This is possible when $f_1T_0R_o \lesssim  r_p$, which approximately means $f_1 \lesssim 10^{-11}$ Hz. Also, we note that the second term involving $r_p$ is less than $20$ and that the last term involving $R_o$ can be dropped compared to the other terms because $2 R_o/c$$\sim$$10^3$ {s}, and~so $2T_0 R_o/c$$\sim$$10^{10}$ {s}$^2$, while the other terms in the bracket are of the order of $10^{14}$ {s}$^2$. Thus, for~$f_1 \lesssim 10^{-11}$ Hz, we can drop these terms, and then the computation of the metric becomes analytically tractable. For~this regime, we write the following:\vspace{-12pt}
\bea
\label{eq_approx_prod_one_spind_phase_2}
  \Phi_{III} (t; \boldsymbol{\vec{\lambda}}) &\approx&  2 \omega_{\text{gw}} t 
+ \hf \omega_{\text{gw}} f_1 \big( 2t^2 + 2T_0 t + T_0^2 \big) \, \no \\
&=& \lambda_0 t + \qtr \lambda_0 \lambda_1 (2 t^2 + 2 tT_0 + T_0^2) \,.
\eea

{To} compute the metric, we simply replace $\Phi_{II}$ in
Equation~(\ref{eq:metric_gen}) with $\Phi_{III}$. Additionally, we find that in this regime, $\lambda_1 T_0 << 1$. For~the integration time, we take $T = T_0$. Carrying out the computations, we obtain the following results:
\be
g_{00} = \frac{1}{12} T_0^2,~~ g_{01} = \frac{1}{12} \lambda_0 T_0^3, ~~ g_{11} = \frac{61}{720} \lambda_0^2 T_0^4 \,.
\ee

{Maximization} over $\lambda_0$ is carried out by minimizing the metric distance with respect to $\lambda_0$. We then obtain $\g_{11} = g_{11} - g_{01}^2/g_{00}$ from $g_{ij}$ as before and find the following:
\be
\sqrt{|\g_{11}|} \sim 0.037 \lambda_0 T_0^2 \, .
\ee

\begin{figure}[H]
    
    \includegraphics[width=0.7\linewidth]{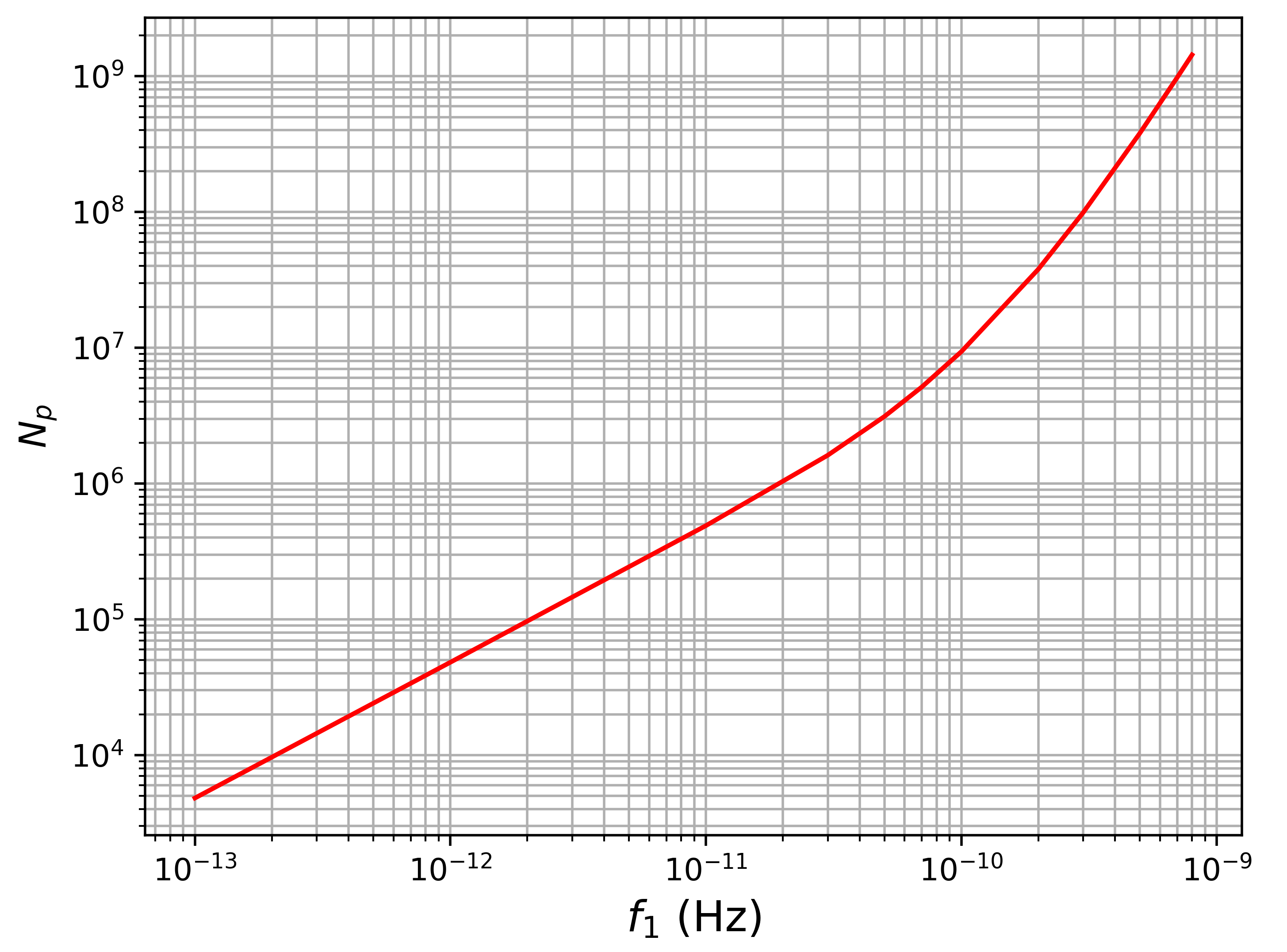}
   
    \caption[Number of patches for the product method (one spin-down case)]{Plot for the number of patches as a function of the maximum spin-down parameter $f_1$ to be searched over. This is for a fixed integration time of $\sim$$6$ months and a maximum GW frequency of $\fgw = 10$ Hz, with the~product signal frequency being $2 \fgw = 20$ Hz.} 
    \label{fig_one_sd_npatch} 
\end{figure}

{The} volume of the parameter space is $V = \sqrt{|\g_{11}|} \Delta \lambda_1$ because~the metric is independent of $\lambda_1$ and $\Delta \lambda_1$ is the range. Since $\Delta \lambda_1$$\sim$$\lambda_1$$\sim$$10^{-11}$ Hz, and~we have  
$\lambda_0 = 4 \pi \fgw$$\sim$$10^2$ rad/s for $\fgw = 10$ Hz, we find that the number of patches is only $\sim$$10^4$ for the spin-down parameter. In this regime, since the~parameters $\a, \del$ decouple from the spin-down, they are essentially uncorrelated. Therefore, we can multiply the number of sky patches in \mbox{Equation~(\ref{eq:area_par})} by the spin-down number (in any case, this is an upper bound). Thus, the~total number of patches considering all the parameters is between $10^5$ and $10^6$. This agrees with the general results obtained numerically, as~is evident from Figure~\ref{fig_one_sd_npatch}. 

The number of patches for the general case must be obtained numerically. We obtain the number of patches as a function of the maximum of the spin-down parameter searched for a given integration time of six months. This is plotted in Figure~\ref{fig_one_sd_npatch}. The~parameter space is now three-dimensional, with~parameters $\a, \del, f_1$ for which we compute the metric $\g_{ij}$. The~volume of the parameter space is obtained by integrating $\sqrt{\det (\gamma)}$ over three parameters $\a, \del, f_1 $. To~determine the number of patches, we use the volume of a single patch as in~\cite{BradyCreightonCutlerSchutz1997}. 
\vspace{-4pt}

As shown in Figure~\ref{fig_one_sd_npatch}, when $f_1 \lesssim 10^{-11}$, the~number of patches scales as $f_1$, while for $f_1 \gtrsim 10^{-11}$, they scale as $f_1^3$. As~a function of the frequency of the GW signal, the~number of patches scales as $\fgw^3$.

\section{The Product~Noise}

We now move to the results part of this paper, where we must compute the sensitivities of various methods and compare them. In~order to compute the sensitivity of our search, we also need to compute the product noise, in addition to~the product signal. The~next subsection is devoted to this aspect.
\par

Since we multiply the data by themselves six months later, the~main component of the noise that affects our analysis is the product noise $N(t) = n(t) n(t + T_0)$. We would like to understand the product noise $N(t)$. The~final statistic will essentially be a sum of such terms, and~due to the generalized central limit theorem~\cite{Fellervol1}, \xdeleted{the product noise} {it} will be \xdeleted{also} Gaussian distributed. 
We assume the following for the detector noise $n(t)$:
\bea 
\langle n(t) \rangle &=& 0 \,, \no \\
\langle \tn (f) \tn^* (f') \rangle &=& \hf S_n (f) \del (f - f') \,. 
\eea

Angular brackets denote the ensemble average. In~the above, we assume that the mean of the noise is zero, it is stationary in the wide sense (WSS), and its second moment in the Fourier domain is described by $S_n(f)$, i.e., the~power spectral density (PSD) of the noise. The~factor of half appears because the PSD defined is one-sided; it is defined only for $f > 0$. The~inverse Fourier transform of the two-sided PSD is the autocorrelation function denoted by $K(t)$, which for stationary noise does not depend on the absolute time. 
\par

It is useful to understand the statistics of $N(t)$ at each $t$. Noises $n(t)$ and $n(t + T_0)$ can be taken to be independent. This implies the following:
\be
\langle N(t) \rangle = \langle n(t) \rangle \langle n(t + T_0) \rangle \equiv 0 \,.
\ee

{Thus} the mean $N(t)$ is also zero. Since $n(t)$ and $n(t + T_0)$ are independent, it follows that $n^2(t)$ and $n^2 (t + T_0)$ are independent~\cite{Dhurandhar2024}. Also, $n(t)$ and $n(t + T_0)$ each have variance $\sigma^2$, then the variance of $N(t)$ is provided by the following:
\be
\label{eq_34}
\langle N^2(t) \rangle = \langle n^2(t) \rangle \langle n^2(t + T_0) \rangle = \sigma^4 \,.
\ee

\begin{figure}[H]
\includegraphics[width=0.45\linewidth]{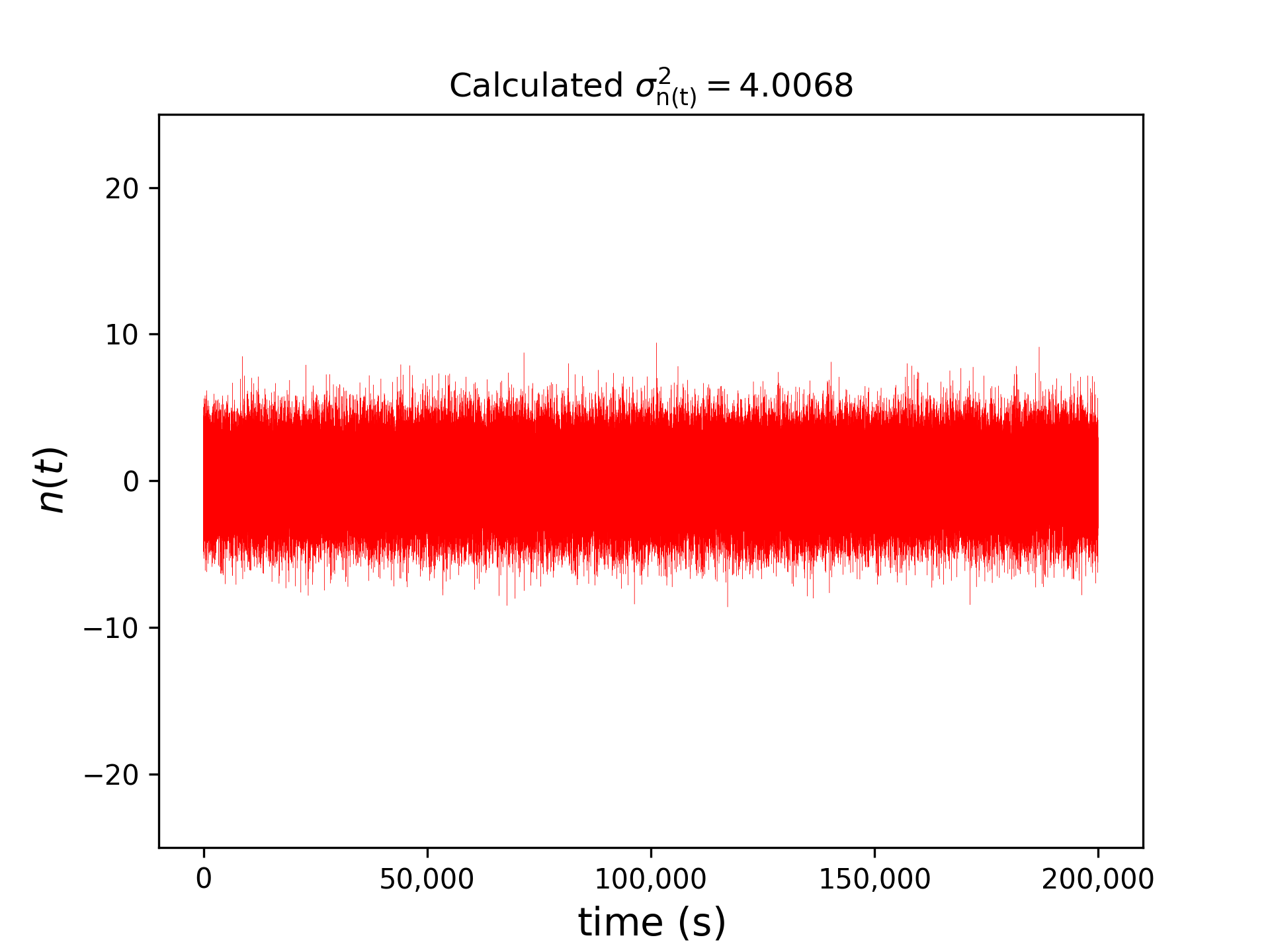}   
\includegraphics[width=0.45\linewidth]{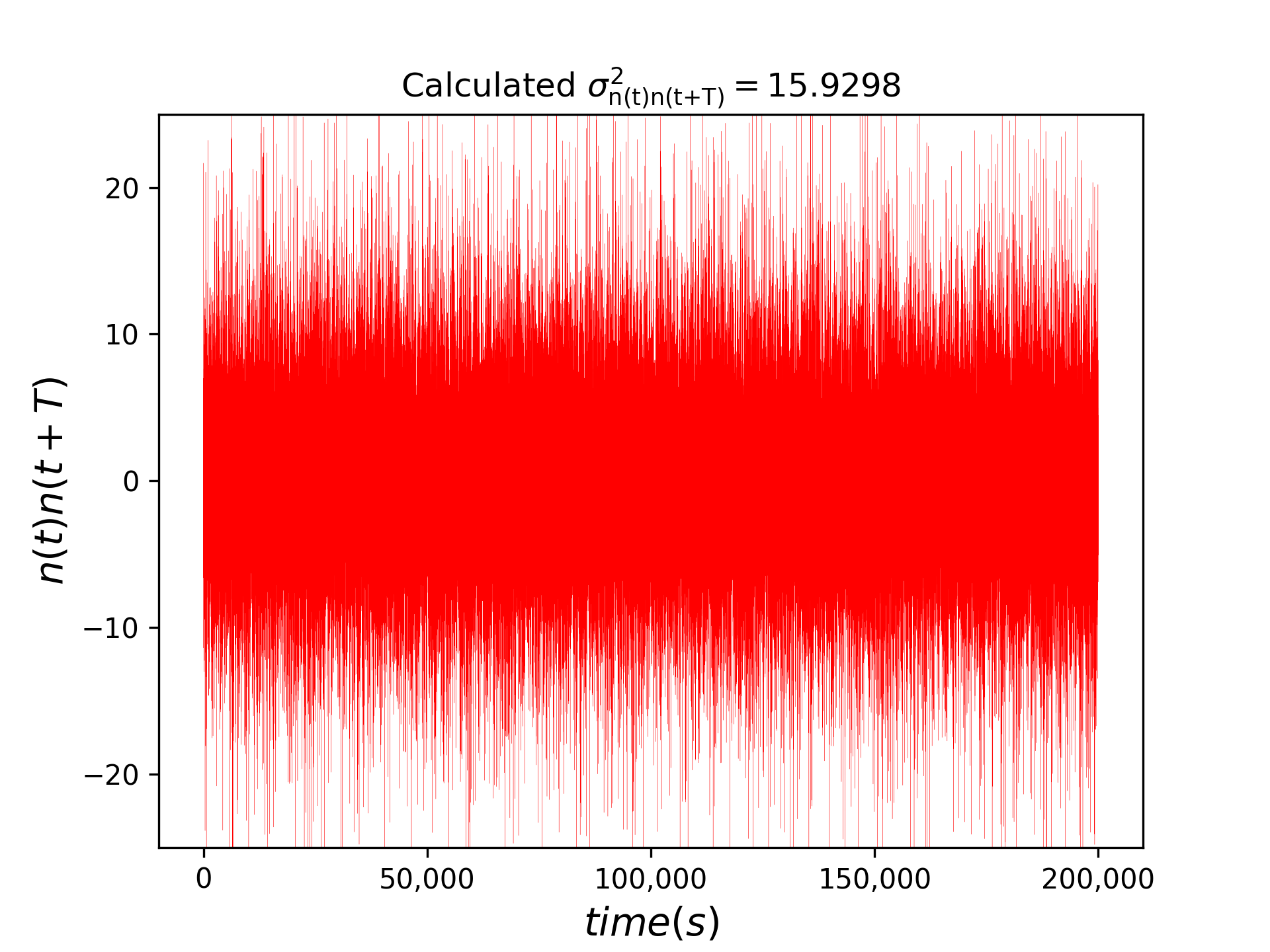}
\caption{The plot on the left shows white noise $n(t)$ with $\sigma^2 = 4$. The~plot on the right shows the  product noise $N(t)$ of two independent Gaussian noise realizations. The~variance of $N(t)$ is $\sigma^4 \approx 16$, which is the square of the variance of $n(t)$. The~data are sampled at 1 Hz, with $t$ plotted on the horizontal~axis.} 
\label{fig:prod_whitenoise} 
\end{figure}

{Also}, it can be shown easily that if $n(t)$ is stationary, then $N(t)$ is also stationary. In~our simulations, we find that for white noise with variance $\sigma^2$, the~product noise is also white with variance $\Sigma^2 = \sigma^4$. This is verified in Figure~\ref{fig:prod_whitenoise} for $\sigma = 2$.
\vspace{-4pt}
In general, assuming independence of noise six months apart and stationarity, we may write the following expression:
\begin{equation}
\langle N(t) N(t + \tau) \rangle = K_1 (\tau) K_2 (\tau) \equiv K(\tau) \,,    
\label{eq:prod_corr}
\end{equation}
where subscripts $1, 2$ denote the autocorrelation functions of the noise six months apart.
\par

We will also assume that noise $n(t)$ is Gaussian with mean zero and variance $\sigma^2$. Then, $N (t)$ has the PDF that is the zeroth-order modified Bessel function of the second kind:
\be
 p(N) = \frac{K_0(|N|/\sigma^2)}{\pi \sigma^2} \,.
 \ee

\begin{figure}[H]
    
    \includegraphics[width=0.6\linewidth]{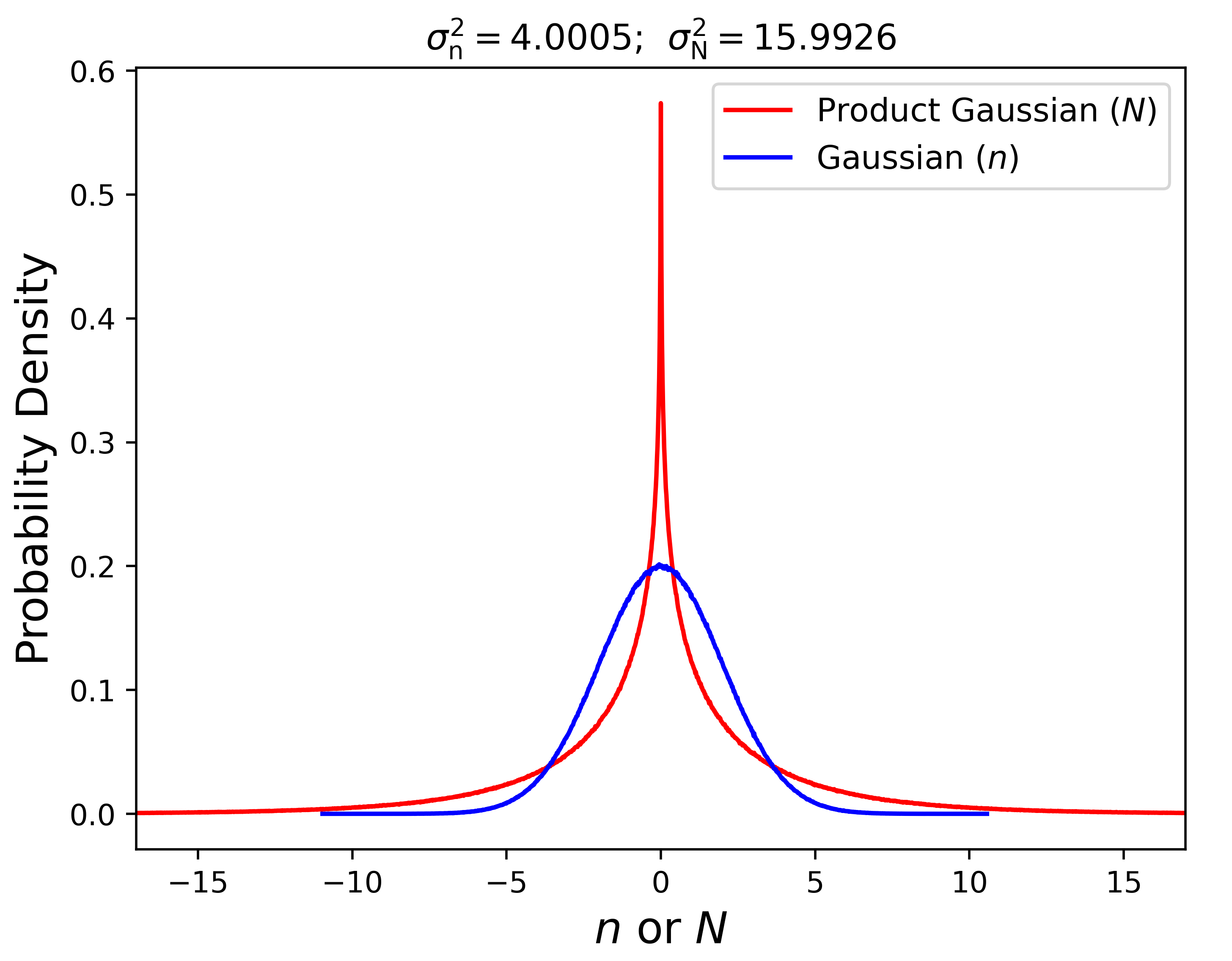}
   
\caption{{Probability} 
 distribution function of two random variables $n$ and $N$. $n$ is a normally distributed random variable, while
$N$ is the product of two identical, independent, and normally distributed random variables. This plot was created using $2\times10^7$ samples of identical non-correlated Gaussian random variables, each with variance
of $\sigma^2 \approx 4$. The~variance of $N$ is $\approx$16, the~square of the variance of~$n$.} 
\label{fig_gauss_prod_gauss}
\end{figure}

Figure~\ref{fig_gauss_prod_gauss} shows the plot of the PDF of $N (t)$ in red, while for comparison, the PDF of $n (t)$, which is Gaussian, is also shown. The~details of this distribution are discussed in \xdeleted{the} Appendix \ref{noise_stats}.
\par

In the Fourier domain, we consider a very narrow band of frequencies around the GW signal because the Doppler modulations change the original frequency of the GW very little- 
 $\Delta \fgw/\fgw \lesssim v/c \sim 10^{-4}$, where $v$ is the orbital velocity of Earth. In~that band, the~noise PSD can be taken to be constant, essentially white. {From Equation~(\ref{eq_34}), it is evident that the  variance of product noise $N$ is the square of the variance of $n$}. Also, the~autocorrelation function {$K(\tau)$} of the product noise is, in~fact, from~Equation~(\ref{eq:prod_corr}), a~product of the autocorrelation functions at time $t$ and time $t + T_0$; thus, {$K(\tau) = K_1 (\tau) K_2 (\tau)$}. Going to the Fourier domain, if~$S_1(f), S_2 (f)$ are the corresponding PSDs, then it is easily shown that the PSD of the product noise is a convolution of the two PSDs. We note that the convolution function usually tends to be a flatter function than individual PSDs because it involves the smoothing operation of integration. For simplicity, we assume that the two noise PSDs are the same, say $S_n (f)$. Assuming a reasonable bandwidth, it follows that, approximately, $4 \langle |{\tilde N} (f)|^2 \rangle \approx S_n (f)^2$, where factor $4$ arises because we have one-sided PSDs. We have carried out extensive simulations with noise curves that support this~assertion.   
\section{Sensitivities and~Thresholds}

\textls[-15]{In this section, we obtain the sensitivities of two methods, the~coherent and product methods, and~compare the two. We assume one spin-down parameter, $f_1$. As~mentioned in the literature, computational resources or power are crucial in obtaining sensitivity: the more computational power, the~more sensitivity the search will have. In~practice, computational power is a fixed resource. Here, we assume that we have a petaflops machine; that is, the~power $P = 10^{15}$ flops (floating-point operations per second). To~fix ideas, first we perform the exercise for $\fgw = 10$ Hz so that the product signal is at $2 \fgw = 20$ Hz.}
\par

{In our analysis, we assume the noise is Gaussian and stationary. However, we observe that for noise in the current detectors, these assumptions are not strictly true, especially at the low frequencies that we are concerned with. Mechanical resonances could interfere with the GW signal and pose a challenge to its detection. One possible way to avoid coupling between
low-frequency mechanical modes and a continuous GW signal present in this frequency region might be to remove the
mechanical modes from the data before implementing our non-linear data processing model. This can be carried out by applying properly selected notch filters to the data. This, of course, would prevent us from detecting a GW signal that would happen to have the same frequency as one of the removed mechanical modes. However, given that the number of resonant frequencies is reasonably small, only a relatively small portion of the frequency band would be removed. This remedy should not compromise the effectiveness of our proposed technique. Secondly, if~our method is seriously considered by experimenters, they could endeavour to remove these noise encumbrances at low frequencies in current or near future detectors.}

\subsection*{{GW} 
 Signal at $\fgw$ = 10 Hz}

First, we analyze the coherent search method and then turn to the product method. We use some of the relevant results from~\cite{BradyCreightonCutlerSchutz1997}. The~number of floating point operations for the search is denoted by $\Nop$. The~main operation in this situation is the fast Fourier transform (FFT). Thus, $\Nop$ is basically the number of floating point operations required to perform FFTs given  the number of sky patches $N_p$. Thus, following~\cite{BradyCreightonCutlerSchutz1997},
\be
\Nop = 6 \fgw T N_p \log_2 (2 \fgw T + 1/2) \,.
\ee

{If} we require these operations to be carried out in time $T$, then fixing the computational power $P$ constrains $T$. In~this case, the~number of patches turns out to be $N_p$$\sim$$3\times10^{11}$ for $f_1 \lesssim 8 \times 10^{-10}$ Hz, which corresponds to $f/{\dot f}$$\sim$$40$ years. We used the results from~\cite{JK3}. We then found that $T$ is constrained to $1.1 \times 10^7$ s.
\par

In the absence of the signal, the~noise power $P_0 (f) = 2 |\tn (f)|^2$ is exponentially distributed (central $\chi^2$ with 2 degrees of freedom) with the PDF:
\be
P (P_0(f)) = \frac{e^{-P_0 (f)/S_n (f)}}{S_n (f)} \,.
\ee

{Then,} the threshold is set by requiring that the probability that any noise event that exceeds the threshold is less than $1 - \a$. Then, the threshold is provided by the following:
\be
\trh \equiv \frac{\rho_c}{S_n (f)} = \ln \frac{\fgw T N_p}{1 - \a} \,.
\label{eq:threshold}
\ee

{If} the power $P_0$ exceeds this threshold, we say that we have detected the signal. We choose $\a = 0.99$. For~these parameters, $\trh$$\sim$$50$. However, if~we require a detection probability of, say, $95 \%$, then the amplitude of the signal should be greater than $\trh$. When a signal is present, the~power follows a non-central $\chi^2$ distribution with 2 degrees of freedom. The~non-central $\chi^2$ distribution with 2 degrees of freedom has a mean of $\mu = \lambda + 2$ and variance of $\sigma^2 = 4 + 4 \lambda$, where $\lambda$ is the non-centrality parameter. When $\lambda >> 1$, the~non-central $\chi^2$ follows a normal distribution with the same mean and variance in the vicinity of the mean~\cite{Wette2012}. It should be noted that, when we switch from the exponential distribution to the $\chi^2$ distribution, the~argument for the $\chi^2$ distribution is twice that of the exponential distribution. Therefore, if~we require the detection probability $P_D$ to be 95 \%, then we must adjust the mean $\mu$ of the signal to be $2 \sigma$ above the threshold $2 \trh$, that is,
\be
\mu = 2 \trh + 2 \sigma = 2 \trh + 2 \times 2 \sqrt{1 + \lambda} \,.
\label{eq:noncentrality_par}
\ee

{This} is an equation for $\lambda$. Solving this equation yields $\lambda$$\sim$$147$. Thus, $\mu = \lambda + 2 \simeq 149$. Then, reverting to the exponential distribution (we must consider $\mu/2$), the~threshold sensitivity with the detection probability $P_D = 0.95$ is as follows:
\be
\hthc  \simeq \sqrt{\frac{(\mu/2) S_h (\fgw)}{T}} \sim \frac{\sqrt{S_h (\fgw)}}{385} \,.
\label{eq:det_prob}
\ee

{The} quantity in the denominator, namely $385$, is called the sensitivity depth, which we denote by $D_{\rm coh}$.
\par

Now, we turn to the product signal, which is at $2 \fgw = 20$ Hz. The~calculation runs essentially on similar lines. We only mention the salient steps. The~noise is now provided by $S_n^2 (2 \fgw)$. We then obtain the following values for the relevant quantities:\mbox{ $N_p$$\sim$$1.4\times10^{9}, ~T$$\sim$$1.5 \times 10^7$ s}. As~before, we assume $1 - \a = 0.01$. Then, from Equation~(\ref{eq:threshold}), we obtain $\trh = \rho_c/S^2_n (f)$$\sim$$50$, where $f \simeq 2 \fgw$. Again, we have a non-central $\chi^2$ distribution with a non-centrality parameter of $\lambda$, which must be evaluated by solving Equation~(\ref{eq:noncentrality_par}). The~result is $\lambda \simeq 147$ and $\mu = \lambda + 2 \simeq 149$, so we obtain a $P_D$ of 95\%.
\par

Now, here is where we differ from the coherent analysis. We must compare $P_0 (f) = 4 |\tH (f)|^2$ (recall that the noise is $4 \langle |{\tilde N} (f) |^2 \rangle$) with~the threshold. This results in a characteristic threshold value of $\hthp$, which is provided by the following equation:
\be
\frac{\hthp^4 T}{S^2_n (f)} = \hf \mu \,.
\ee

{Solving} for $\hthp$, we obtain the following:
\be
\hthp = \frac{\sqrt{S_n (2 \fgw)}}{D_{\rm prod}} \,,
\ee
where
\be
D_{\rm prod} = \left[ \frac{2 T}{\mu} \right]^\qtr  \sim 21.8 \,.
\ee

{Here}, $D_{\rm prod}$ is the depth of the product~search.

\begin{figure}[H]
    
    \includegraphics[width=0.6\linewidth]{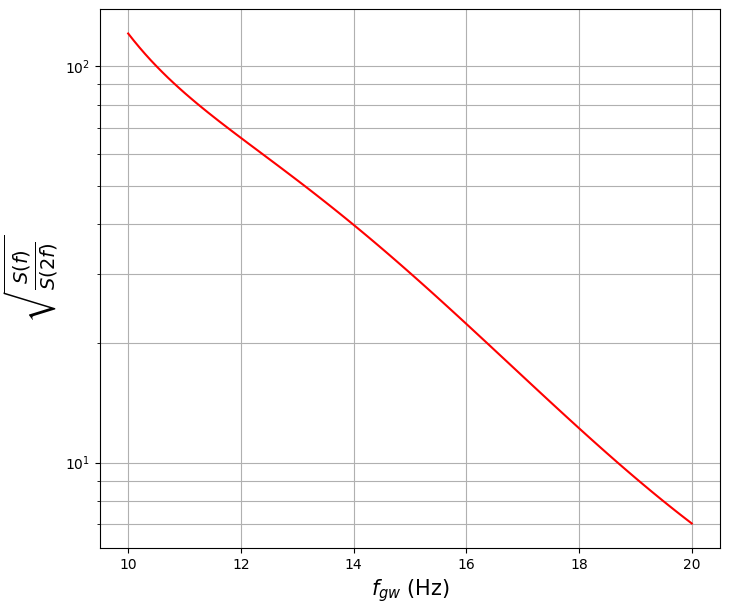}
   
\caption{{The} 
 ratio of the square root of the PSDs: $\sqrt{S_n (f)/S_n (2f)}$ is plotted for the O4 observation run. This plot was generated using a smoothed power spectral density (PSD) estimate from the O4~run.} 
\label{fig_psd_ratio}
\end{figure}

There is a factor of about $385/21.8$$\sim$$18$ between the sensitivity depths. However, we emphasize that this factor is hugely offset by the ratio of the PSDs at $10$ and $20$ Hz. This is the power of our method. This factor, $\sqrt{S_n (\fgw)/S_n (2\fgw)}$, is plotted in Figure~\ref{fig_psd_ratio} for the O4 noise PSD. At~$10$ Hz, as~seen in the figure, this factor is more than $100$, overwhelmingly compensating for the ratio of the sensitivity depths.
\par
\begin{figure}[H]
    
    \includegraphics[width=0.6\linewidth]{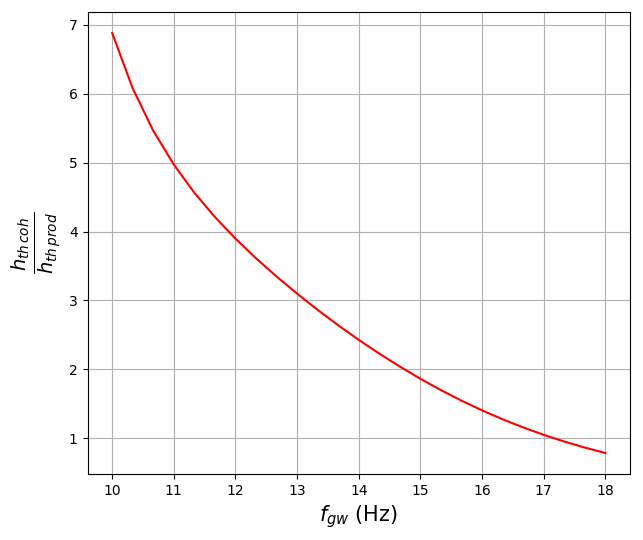}
   
    \caption{The ratio of sensitivities of the product method and the coherent method plotted for the O4 run PSD. Note that the frequency of the product signal is $2 \fgw$. We see that the curve crosses unity at $\sim$$17$ Hz, which means that the product method performs better than the coherent method at frequencies less than $\sim$$17$ Hz.} 
    \label{fig_rel_sen}
\end{figure}

\begin{figure}[H]
    
    \includegraphics[width=0.6\linewidth]{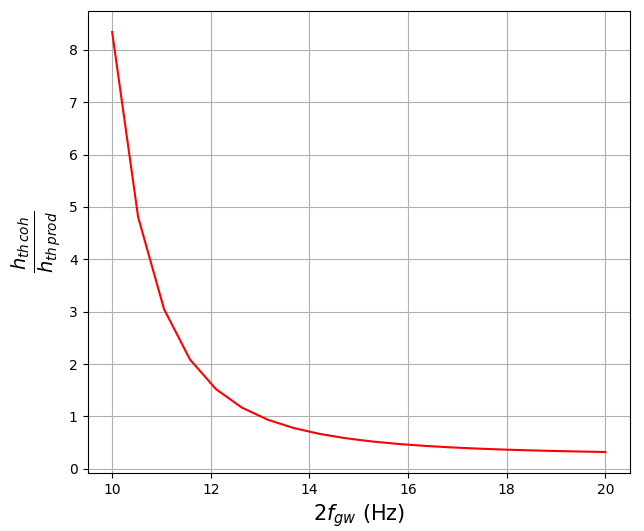}
   
    \caption{\textls[-15]{Ratio of sensitivities of the product method and the coherent method for the design sensitivity curve of advanced LIGO. The~product method performs better for frequencies of $2 \fgw \lesssim 13$ Hz.}} 
    \label{fig_rel_sen_design}
\end{figure}

We plot the ratio of the sensitivities, i.e., the relative sensitivity, of signals at a given GW frequency in Figure~\ref{fig_rel_sen}, where we used the O4 noise PSD. We see that our method performs better than the fully coherent method~\cite{BradyCreighton1998} for frequencies lower than $\lesssim$17 Hz. In~addition, we evaluated the relative sensitivities for the advanced LIGO design sensitivity curve. The~product method performs better than the coherent method for $2f_{gw} \lesssim 13$ Hz. This is shown in Figure~\ref{fig_rel_sen_design}.
\vspace{-4pt}
\unskip

{It should be noted that the computing power of petaflops enforces a constraint on the time of integration for the coherent search. If,~however, we increase the computing resources, we could carry out the coherent search for the full six months or even longer. For~six months of data, we would require computing power of $20$ petaflops to analyze the data in an even time. This increase results essentially from the increase in sky patches, a~factor of $\sim$$20$ (the increase in FFT length is compensated by the corresponding time available for the data analysis, as is usually assumed~\cite{BradyCreightonCutlerSchutz1997}). The~sensitivity of the coherent search $\propto \sqrt{T}$ would then increase by about $20 \%$, which would not affect our conclusions. Even if we extend the search to a year's worth of data, this would not compensate for the loss in sensitivity at lower frequencies.}
\par

We remark that we have tacitly assumed that we have only one peak, but~in fact, several peaks appear in the matched filtering operation. Our analysis applies to the dominant peak, and~the appearance of more than one peak makes little difference to our~results.

\section{From Circular to~Elliptical}
\label{el_orb}
The analysis presented so far assumes the trajectory of Earth to be perfectly circular. As~we shall now show, effects due to eccentricity $e$ would significantly degrade our results if not taken into account. In~what follows, we first quantify the effects of the eccentricity $e$ of Earth's trajectory around the SSB and~then proceed to show how to compensate for it so that the results derived earlier for a circular trajectory remain~valid. 

We describe the trajectory of the center of the Earth around the Sun as elliptical. The~gravitational perturbations caused by other planets (mainly Jupiter and Saturn) can be disregarded over a six-month time scale. If~we focus on Earth's velocities
at the apogee and perigee (exactly six months apart), we recognize that at these two points, their signs are exactly opposite, but~their magnitudes differ the most. Their percentage difference is in fact equal to $2e \simeq 3.4 \times 10^{-2}$. This means that the corresponding frequency shift, which depends on the location of the source and is proportional to the GW frequency $f_0$, could be as large as $3.2 \times 10^{-4}$ Hz for a signal with a frequency of $f_0 = 100$ Hz. At~10 Hz, this effect is smaller by a factor of 10. This means that if ignored, the~eccentricity would degrade our results derived for a circular~orbit.

We first note that at each moment in time, it is still possible to identify two points on Earth's trajectory that are diametrically opposite with respect to the center of the ellipse. However, it is clear that their separation in time is no longer equal to six months. As~we shall show, the~time shift, $T_E$, allowing the removal of Doppler modulation in the product dataset, is itself a function of time $t$.

To this end, let us take the origin of our coordinate system to be the center of Earth's trajectory around the Sun and define it with ($x, y$) two orthogonal axes in the plane of the ecliptic along the semi-major and semi-minor axes, respectively. In~this coordinate system, Earth's trajectory around the Sun can be described mathematically by the
following expressions~\cite{Landau1976Mechanics}:
\begin{eqnarray}
t & = & \sqrt{ \frac{\mu a^3}{\alpha} }
      \left[\xi - e \sin(\xi)\right] \ ,
\label{t}
\\
  x & = & a \cos(\xi) \ ,
\label{x}
\\
  y & = & a \sqrt{1 - e^2} \sin(\xi) \ .
\label{y}
\end{eqnarray}

{In} Equation~(\ref{t}), $t$ corresponds to the time elapsed since Earth was in perihelion and is expressed in parametric form through the angle $\xi$, which represents the angular location of Earth from
perihelion. The~remaining physical quantities appearing in the above equations are the semi-major axis ($a$), the~reduced mass of the
Sun--Earth system ($\mu$), the~eccentricity ($e$), and~$\alpha \equiv G m_1 m_2$, with ($m_1, m_2$) being the masses of the two bodies and being $G$ the gravitational~constant.

In this chosen coordinate system, each point on the trajectory, represented by a vector
${\vec r}_E (\xi)$, has a diametrically opposite point,
${\vec r}_E (\xi + \pi)$, such that
${\vec r}_E (\xi) + {\vec r}_E (\xi + \pi) = 0$. Since $t$ is a non-linear function of $\xi$, we can identify time $T_E$ such that
${\vec r}_E (\xi + \pi) = {\vec r}_E (t + T_E)$. This can be carried out by
noticing that Equation~(\ref{t}) above implies the following expression:
\begin{equation}
  t + T_E = \sqrt{ \frac{\mu a^3}{\alpha} }
  \left[\xi + \pi - e \sin(\xi + \pi)\right] \ .
  \label{tT}
\end{equation}

{After} taking the difference between Equations~(\ref{tT}) and~(\ref{t}),  we obtain the following expression for $T_E$:
\begin{equation}
T_E = \sqrt{ \frac{\mu a^3}{\alpha} }
  \left[\pi + 2 e \sin(\xi)\right] \ .
  \label{TE}
\end{equation}

{Since} both $t$ and $T_E$ are functions of $\xi$, it follows that $T_E$ can
be considered a function of $t$. For~any given time $t$,
Equation~(\ref{t}) can be solved numerically for $\xi (t)$, and~$T_E(t)$ can be obtained from the above equation. Similar to what happens with a circular trajectory, here, the product of interferometric measurements made at time $t$ and $t +
T_E(t)$ also results in the exact cancellation of the Doppler modulation due to the movement of Earth around the Sun. The~expression of the phase of the product signal at time $t$ and $t + T_E(t)$ is now identical to the expression provided in Section~\ref{SecIII}, but~with $T_E$ as a known function of time $t$.

If we now substitute the values for the parameters entering the
expression for $T_E (t)$ and~expand the dependence of $t$ on $\xi$,
we can rewrite $T_E(t)$ in the following form:
\begin{equation}
T_E (t) = T_0 [1 + \frac{2 e}{\pi} \sin(\xi) ] \simeq
T_0 [1 + \frac{2 e}{\pi} \sin(\frac{\pi t}{T_0})] \ ,
  \label{TEE}
\end{equation}
where $T_0 = 6 \ {\rm months}$. The~above equation shows that $T_E(t)$ is a slowly changing function of time with a period equal to $T_0$, and, as~expected, it differs due to eccentricity.

\section{Summary and~Conclusions}
\label{SecV}

We have quantified the advantages of a recently proposed non-linear data processing technique~\cite{TintoCW2021} to search for continuous GW signals in the data measured by Earth-based detectors. This technique takes advantage of the symmetry of the motion around the Sun of an Earth-bound gravitational wave interferometer by multiplying the measured data time series with a half-year shifted copy of it. This technique offers two main advantages. First, the~main Doppler phase modulation of a monochromatic GW signal is exactly removed; second, the~frequency of the signal in the product data is equal to twice the GW signal frequency. The~first effect significantly reduces the size of the signal parameter space over which the search is performed. As~we have shown, this implies that our proposed coherent method can be implemented over a year-long data segment and requires a processing time comparable to the data acquisition time with currently available computers. The~second is that it becomes advantageous at low frequencies. In~fact, we have found that our technique is capable of achieving sensitivity that is better than that of coherent and possibly other non-coherent methods in this part of the accessible frequency band. Also, our method turns out to be much faster than the coherent method, typically by a factor of $10^4$.

The technique we have discussed in this article provides an additional tool for processing the data generated by Earth and future space-based interferometers by taking advantage of the symmetry of the motion of the detectors around the Sun. In a forthcoming article, we will investigate how this approach can be used to effectively disentangle the superposition of hundreds of millions of signals from white dwarf--white dwarf binary systems expected to affect the sensitivity of space-based interferometers such as the Laser Interferometer Space Antenna (LISA) and the TaiJi missions~\cite{LISA2017,Taiji}.
\vspace{6pt}

\authorcontributions{The authors of this article contributed equally
  to the work~reported. Conceptualization, H.R., S.D. and M.T.; methodology,  H.R., S.D. and M.T.; software,  H.R.; validation,  H.R., S.D. and M.T.; formal analysis,  H.R., S.D. and M.T.; investigation,  H.R., S.D. and M.T.; resources,  H.R., S.D. and M.T.; data curation,  H.R., S.D. and M.T.; writing---original draft preparation,  H.R., S.D. and M.T.; writing---review and editing, H.R., S.D. and M.T.; visualization, H.R.; supervision, S.D. and M.T.; project administration, S.D. and M.T.; funding acquisition,  H.R., S.D. and M.T. All authors have read and agreed to the published version of the~manuscript.}

\funding{{For} 
 H.R. and S.D., this research was self-funded. For~ M.T., this research was funded by the Polish National Science Center, Grant No.  2023/49/B/ST9/02777.}

\dataavailability{Data are contained within the article. 
}

\acknowledgments{{We} 
 thank our institutions (IISER, IUCAA, and INPE) for their kind hospitality while this work was carried out.}

\conflictsofinterest{The authors declare no conflicts of interest. The~ funders had no role in the design of the study; in the collection, analysis, or~interpretation of data; in the writing of the manuscript; or in the decision to publish the~results.}


\abbreviations{Abbreviations}{
The following abbreviations are used in this manuscript:\\
 
\noindent
\begin{tabular}{@{}ll}
GW & Gravitational Wave\\
SSB & Solar System Barycenter\\
FFT & Fast Fourier Transform
\end{tabular}
}

\vspace{6pt}

\appendixtitles{yes} 
\appendixstart
\appendix
\section{The Amplitudes Describing the Product Signal in Case~I}
\label{Prod_Sig}

The amplitudes of various terms in Equation (\ref{eq:gnrl_signal}) are provided in terms of the coefficients, which we present in four steps. First, we define the following functions:
\begin{equation}
    \begin{aligned}
        \mathcal{C}_1(i, \delta, \theta_l, \gamma) &= 12(1 + \cos^2 i)\cos^2 \delta \cos^2 \theta_l \sin 2\gamma, \\
        \mathcal{C}_2(i, \delta, \theta_l, \gamma) &= (\cos 2\theta_l - 3)\sin 2\gamma(\cos 2\delta - 3)(1 + \cos^2 i), \\
        \mathcal{C}_3(i, \delta, \theta_l, \gamma) &= -4 \sin 2\delta \sin 2\gamma \sin 2\theta_l (1 + \cos^2 i), \\
        \mathcal{C}_4(i, \delta, \theta_l, \gamma) &= -8 \sin 2\delta \cos 2\gamma \cos \theta_l (1 + \cos^2 i), \\
        \mathcal{C}_5(i, \delta, \theta_l, \gamma) &= -4 \cos 2\gamma (\cos 2\delta - 3)\sin \theta_l (1 + \cos^2 i), \\
        \mathcal{S}_1(i, \delta, \theta_l, \gamma) &= -32 \cos^2 i \cos 2\gamma \cos \delta \cos \theta_l, \\
        \mathcal{S}_2(i, \delta, \theta_l, \gamma) &= 32 \cos^2 i \cos 2\gamma \sin \delta \sin \theta_l, \\
        \mathcal{S}_3(i, \delta, \theta_l, \gamma) &= 16 \cos^2 i \sin 2\gamma (2 \cos \delta \sin 2\theta_l), \\
        \mathcal{S}_4(i, \delta, \theta_l, \gamma) &= 8 \cos^2 i (\cos 2\theta_l - 3) \sin \delta \sin 2\gamma.
    \end{aligned}
\end{equation}

The next set of functions is as follows:
\begin{equation}
    \begin{aligned}
        \mathcal{G}_{c0}(i, \delta, \theta_l, \gamma) &= \C_1^2 + \frac{1}{2} \{ \C_2^2 + \C_5^2 + \S_1^2 + \S_3^2 - \C_3^2 - \C_4^2 - \S_2^2 - \S_4^2\}, \\
        \mathcal{G}_{c1}(i, \delta, \theta_l, \gamma) &= 2\C_1\C_2 + \frac{1}{2} \{ \S_1^2 + \C_4^2 - \S_3^2 - \C_3^2\}, \\
        \mathcal{G}_{c2}(i, \delta, \theta_l, \gamma) &= \C_3 \C_4 - \S_1\S_3 - 2\C_1\C_5, \\
        \mathcal{G}_{c3}(i, \delta, \theta_l, \gamma) &= \frac{1}{2}(\C_2^2 + \S_4^2  - \S_2^2 - \C_5^2), \\
        \mathcal{G}_{c4}(i, \delta, \theta_l, \gamma) &= \S_2\S_4 - \C_2\C_5. \\
    \end{aligned}
\end{equation}

\begin{equation*}
    \begin{aligned}
        \mathcal{G}_{s1}(i, \delta, \theta_l, \gamma) &= 2\C_1\S_2 - \C_3\S_1 + \C_4\S_3, \\
        \mathcal{G}_{s2}(i, \delta, \theta_l, \gamma) &= \C_4 \S_1 + \C_3\S_3 - 2\C_1\S_4, \\
        \mathcal{G}_{s3}(i, \delta, \theta_l, \gamma) &= \C_2\S_2 - \C_5\S_4, \\
        \mathcal{G}_{s4}(i, \delta, \theta_l, \gamma) &= -\C_5\S_4 - \C_2\S_4. \\
    \end{aligned}
\end{equation*}

Further, we define the following set of functions:
\begin{equation}
    \begin{aligned}
    A_0 &= 2K\mathcal{G}_{c0}, \\
    B_0 &= 0\\
    A_{1c\pm} &= K(\mathcal{G}_{c1} \pm \mathcal{G}_{s2}), \\
    A_{1s\pm} &= K(\mathcal{G}_{c2} \mp \mathcal{G}_{s1}), \\
    B_{1c\pm} &= K(\mathcal{G}_{s1} \mp \mathcal{G}_{c2}), \\
    B_{1s\pm} &= K(\mathcal{G}_{s2} \pm \mathcal{G}_{c1}), \\
    A_{2c\pm} &= K(\mathcal{G}_{c3} \pm \mathcal{G}_{s4}), \\
    A_{2s\pm} &= K(\mathcal{G}_{c4} \mp \mathcal{G}_{s3}), \\
    B_{2c\pm} &= K(\mathcal{G}_{s3} \mp \mathcal{G}_{c4}), \\
    B_{2s\pm} &= K(\mathcal{G}_{s4} \pm \mathcal{G}_{c3}), \\ 
    \end{aligned}
\end{equation}
where
\begin{equation}
    K = \frac{h_0^2}{4096}.
\end{equation}

Finally, we define the coefficients in Equation~(\ref{eq:gnrl_signal}) as follows:
\begin{equation}
    \begin{aligned}
    &A_0=2K\mathcal{G}_{c0},\\
    &B_0=0,\\
    &A_{\pm1} = \{ A_{1c\pm} \cos2\alpha + A_{1s\pm} \sin2\alpha \}, \\
    &B_{\pm1} = \{ B_{1c\pm} \cos2\alpha + B_{1s\pm} \sin2\alpha \}, \\
    &A_{\pm2} = \{ A_{2c\pm} \cos4\alpha + A_{2s\pm} \sin4\alpha \}, \\
    &B_{\pm2} = \{ B_{2c\pm} \cos4\alpha + B_{2s\pm} \sin4\alpha \}. \\
    \end{aligned}
\end{equation}

{The} coefficient $B_0$ is zero because of the choice of coordinates and initial phases of the rotation of the~Earth.

\section[\appendixname~\thesection]{The Finite-Time Delta Function}
\label{append_deltafn}

The usual delta function can be described in terms of a Fourier integral:
\begin{equation}
\del (\om) ~=~ \int_{- \infty}^{\infty} ~e^{-i \om t} ~dt \,.
\end{equation}

{Since} our data length is finite, say, of~duration $T$, the~above
definition needs to be modified.  Accordingly, we define the finite
time delta function $\del_T (\om)$ by the integral:
\begin{equation}
\del_T (\om) ~=~ \int_{- T/2}^{T/2} ~e^{-i \om t} ~dt \,.
\end{equation}

{For} convenience, we take the integral symmetrically at about $t =
0$. The~integral can be easily performed and yields the following result:
\begin{equation}
\del_T (\om) ~=~ T ~\frac{\sin \om T/2}{\om T/2} \equiv T ~{\rm sinc} ~ \om T/2 \,.
\end{equation}

{The} $\del_T (\om)$ is essentially a sinc function. The~following properties hold: 
\begin{eqnarray}
\delta_T (0) &=& T \,, \no \\
\frac{1}{2 \pi}\int_{- \infty}^{\infty} ~\del_T (\om)~d \om &=& 1 \,, \no \\
\frac{1}{2 \pi}\int_{- \infty}^{\infty} ~ |\del_T (\om)|^2~d \om &=& T \, .
\end{eqnarray}

{This} means $\| \del_T (\om) \|^2 = T$, where we have used the $L^2$
norm. One may divide $\del_T (\om)$ by $\sqrt{T}$ to obtain a normalized
finite time delta function ${\hat \del}_T (\om)$.  We have
\begin{equation}
{\hat \del}_T (\om) = \frac{1}{\sqrt{T}} \del_T (\om) \,.
\end{equation}

{Then}, $\| {\hat \del}_T (\om) \| = 1$.
\par

The finite-time delta functions are essentially sinc functions. The~fractional interference between $\del_T (\om)$ and
$\del_T (\om + \Delta \om)$ depends on both $\Delta \om$ and $T$. It is
in essence the uncertainty principle. Since we have the inequality
$| \sin x/x | \leq 1/x$, this provides an upper bound for the sinc
function.  Thus, considering the first two terms in
Equation~(\ref{eq:mf_out}), we see that the fractional interference is
determined by the quantity $\ome T$ and, therefore, the~upper bound is
$(\ome T)^{-1}$. So, interference can be at most $\sim$$5$\% when
$T$ is about 3 days; that is, $\ome T \approx 20$, and~it falls to less
than $1$\% when $T$$\sim$$17$ days. The~figure below shows the
interference of sinc functions centered at $x =10$ and $x = 30$. The~frequency is~unity.
\vspace{-4pt}
\begin{figure}[H]

\includegraphics[width=0.6 \textwidth]{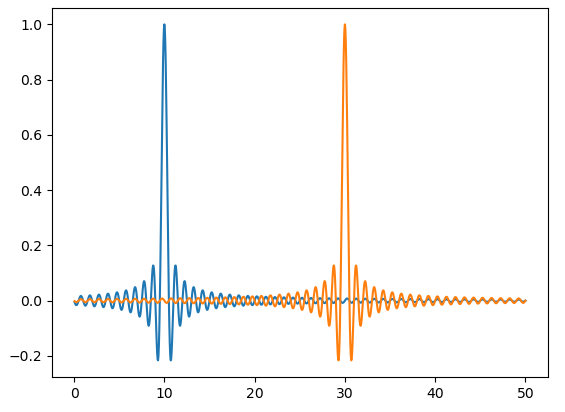}

\caption{{Interference} 
 of two sinc functions centered at points
  $x = 10$ and $x = 30$ with frequency $f = 1$. Here, we have
  $\Delta x = 20$ and $\om = 2 \pi$. It is evident from the figure
  that the interference is less than $5 \%$. For~3 days,
  $\Delta \om = \ome T \approx 20$.}
\label{fig:sinc_functions}
\end{figure}
\unskip 

\section[\appendixname~\thesection]{The Statistics of the Product Noise}
\label{noise_stats}

We consider two independent Gaussian random variables, $X$ and $Y$, with
a mean of zero and variance of $\sigma^2$. Our aim is first to compute the PDF
of $Z = XY$. One method is to define an auxiliary variable
$W = \hf (X^2 - Y^2)$ and compute the joint PDF of $W, Z$ and then marginalize it over $W$. First, let us define $\sigma = 1$ to avoid clutter. We will insert it later on. Then, we have the~following:
\begin{equation} 
P(x, y) = \frac{1}{2 \pi}
e^{- \hf (x^2 + y^2)} \,.  
\end{equation}

{Also}, $4 (w^2 + z^2) = (x^2 + y^2)^2$. Inserting the Jacobian
$2 \sqrt{w^2 + z^2}$, we obtain some simple algebraic
manipulations for $z > 0$ and marginalization over $w$:

\begin{equation}
p_Z (z) = 2 \times \frac{1}{2 \pi} \int_0^\infty d \theta~ e^{-z \cosh \theta} = \frac{K_0 (z)}{\pi}\,.  
\end{equation}

{A} factor of $2$ is included to obtain the correct PDF because~the
$(x, y)$ plane maps {\it twice} into the $(w, z)$ plane, where the points
$(x, y)$ and $(-x, -y)$ map to the same point $(w, z)$. Or, from~another point of view, the~half-plane $x \geq 0$ maps out the full $(w, z)$ plane.

Since $p_Z(z)$ is a symmetric function of approximately $z = 0$, we have
\begin{equation}
p_Z(z) = \frac{K_0(|z|)}{\pi}.
\end{equation}

Scaling by $\sigma$ can be easily achieved by scaling the individual variables $X, Y$, and the final result is the following:
\begin{equation}
p_Z(z) = \frac{K_0(|z|/\sigma^2)}{\pi \sigma^2}.
\end{equation}

{In} order to obtain the variance of $Z$, we use the following formula~\cite{abramowitz_stegun}:
\begin{equation}
\int_0^\infty dz~z^\mu K_0(z) dz = 2^{\mu - 1} \left[\Gamma \left(\frac{\mu + 1}{2} \right) \right]^2 \,.
\end{equation}

{Setting} $\mu = 2$ and the scaling variables, we find that the variance of $Z$ is $\sigma^4$. For~an example, refer to Figure~\ref{fig_gauss_prod_gauss}.    

\vspace{8pt} 


\begin{adjustwidth}{-\extralength}{0cm}

\printendnotes[custom]

\reftitle{References}



\PublishersNote{}

\end{adjustwidth}

\end{document}